# Anomalous Hall Effect in Fe/Gd Bilayers

W. J. Xu[1], B. Zhang[2], Z. X. Liu[1], Z. Wang[1], W. Li[1], Z. B. Wu[3], R. H. Yu[4]

and X. X. Zhang[2*]

[1] Dept. of Phys. and Institute of Nanoscience & Technology, The Hong Kong University of Science and Technology (HKUST), Clear Water Bay, Kowloon, Hong Kong, P. R. China

[2] Image-characterization Core Lab, Research and Development, 4700 King Abdullah University of Science and Technology (KAUST), Thuwal 23900-6900, Kingdom of Saudi Arabia

[3] Laboratory of Advanced Materials, Department of Materials Science and Engineering, Tsinghua University, Beijing 100084, P.R. China

[4] School of Materials Science and Engineering, Beihang University, Beijing 100191, P. R. China

**Abstract.**

Non-monotonic dependence of anomalous Hall resistivity on temperature and magnetization, including a sign change, was observed in Fe/Gd bilayers. To understand the intriguing observations, we fabricated the Fe/Gd bilayers and single layers of Fe and Gd simultaneously. The temperature and field dependences of longitudinal resistivity, Hall resistivity and magnetization in these films have also been carefully measured. The analysis of these data reveals that these intriguing features are due to the opposite signs of Hall resistivity/or spin polarization and different Curie temperatures of Fe and Gd single-layer films.

In addition to the ordinary Hall effect ($R_O B$), a magnetization (M) -dependent contribution to the Hall resistivity ($\rho_{xy}$) is commonly observed in ferromagnetic materials,

$$\rho_{xy} = R_O B + R_S 4\pi M = R_O \left[ H + 4\pi M (1-N) \right] + R_S 4\pi M, \qquad (1)$$

where H is the applied magnetic field and N is the demagnetization factor [1]. Since the coefficient, $R_S$, is generally much larger (at least 10 times) than the ordinary constant, $R_O$, the anomalous Hall effect ($AHE = R_S 4\pi M$) has been useful in the investigation and characterization of itinerant electron ferromagnets [1]. In the last few years, AHE has been used in studies of the mechanism of ferromagnetism in diluted magnetic semiconductors (DSM), the materials with the best potential for spintronic devices [2,3,4]. Although AHE was discovered more than a century ago [1], its mechanism is not yet well understood. The first theoretical investigation by Karplus and Luttinger [5] showed that spin-orbit coupling in Bloch bands in a perfect ferromagnetic crystal should be the origin, known as the *intrinsic* effect of AHE. Later, the extrinsic scattering mechanisms of skew-scattering (SS) and side-jumps (SJ) were proposed by Smit [6] and Berger [7], respectively. The skew-scattering model predicts that there is a linear dependence of anomalous Hall resistivity ($\rho_{AHE}$) on the longitudinal resistivity ($\rho_{xx}$), i.e.

$$\rho_{AHE} \sim \rho_{xx}; \qquad (2)$$

whereas the side-jump model predicts a quadratic dependence,

$$\rho_{AHE} \sim \rho_{xx}^2. \qquad (3)$$

Since linear and/or quadratic dependence of AHE on longitudinal resistivity has been commonly observed experimentally, it has been suggested that AHE in ferromagnetic materials is an extrinsic effect and is due solely to carrier scatterings on impurities. However, a quantitative comparison between experiments and theory is not possible due to the lack of information on the scattering potential of real materials [8,9].

Interestingly, the intrinsic mechanism proposed by Karplus and Luttinger [5] was recently re-examined [2,8,10,11,12]. It is found that the anomalous Hall conductivity can indeed be caused by the Berry curvatures in momentum space [8,12], where the Berry curvature is defined by [8]

$$\Omega_n(\mathbf{k}) = -\mathrm{Im}\langle \nabla_{\mathbf{k}} u_{n\mathbf{k}} | \times | \nabla_{\mathbf{k}} u_{n\mathbf{k}} \rangle, \qquad (4)$$

with $u_{n\mathbf{k}}$ the Bloch wave function in the $n$th band. The anomalous Hall conductivity can then be written as the sum of the Berry curvatures over the occupied Bloch states. The intrinsic anomalous Hall conductivity has been evaluated numerically in bcc iron [8], ferromagnetic semiconductors [2] and oxides [11]. It was found that the intrinsic anomalous Hall conductivity (AHC) can be quite large, and a quantitative agreement between theory and experiment is now possible.

AHE in colossal magnetoresistance manganites [11,13], half metallic $CrO_2$ films [14] and Gd single crystals [15] has been related to the Berry phase effect. A complicated temperature dependence of anomalous Hall resistivity (AH resistivity) in the itinerant ferromagnet $SrRuO_3$ was reported by Fang et al. [10]. For example, $\rho_{AHE}$ is a non-monotonic function of temperature and its sign even changes. Fang et al. interpreted this behavior with the Berry phase effect in the crystal momentum space. To test this model, Kats et al. [16] repeated the measurements by varying the magnetic field at various temperatures in the vicinity of the temperature at which $\rho_{AHE}$ changes its sign. They claimed that the results seemed to contradict the Berry phase mechanism.

In this paper, we describe our observation of the non-monotonic dependence of AH resistivity on the temperature and magnetization, including a sign change around room temperature, in Fe/Gd bilayers, Rather than the more complex description by intrinsic Hall effect, however, all the features observed can be interpreted within a simple model developed for Fe/Cu bilayers [17] by considering the differing spin polarization in Fe and Gd (the signs of AHE in Fe and Gd are opposite).

Three samples, Fe (200 nm) single layers, Gd (200 nm) single layers and Fe (200 nm)/Gd (200 nm) bilayers were fabricated simultaneously in one run using multi-target, magnetron sputtering on glass and single crystal quartz disk (5 mm in diameter and 0.1 mm thick) substrates. The base pressure and argon pressure for sputtering were $2\times10^{-7}$ Torr and $4\times10^{-3}$ Torr, respectively. To perform Hall and resistance measurements simultaneously and to guarantee that the data for electrical transport (Hall and longitudinal resistivity) measurements were obtained from the same sample, masks were used to make patterned samples. To protect the films from oxidation, all the samples were covered with a thin $SiO_2$ layer. The film thickness was controlled by the sputtering power and deposition time and then measured with a Veeco-Dektak surface profiler. Patterned films on glass substrates were used for the resistivity and Hall measurements with a Quantum Design Physical Property Measurement System. Magnetization data were obtained by measuring the films deposited on the quartz disks with a Quantum Design Magnetic Property Measurement System. The residual longitudinal offset in measured

Hall voltage was excluded using its different symmetry against the magnetic field, $H$, and the anomalous part ($\rho_{xy}^{A}$) of the Hall resistivity was extracted according to eq. (1).

The Hall resistivity, $\rho_{xy}$, for all samples was measured in a magnetic field −5 T to 5 T magnetic field and at temperatures ranging from 5 K to 350 K. From the field-dependent anomalous Hall resistivity curves, we take the 5 T as the saturated anomalous Hall resistivity. Shown in fig. 1 is the temperature dependent anomalous Hall resistivity for the Fe (200nm)/Gd (200nm) bilayer. As the temperature increases from 5 K, the magnitude of the negative AHE increases slightly from 0.6 μΩcm and reaches a smooth maximum ~0.7 μΩcm at about 150K. With further increasing temperature, the magnitude of AHE decreases almost linearly from 200 K and vanishes at 280 K. Above 280 K, AHE becomes positive and increases as temperature increases. The non-monotonic temperature dependence of AH resistivity including a sign change at 280 K is the most intriguing feature in fig. 1. A monotonic temperature dependence of AHE resistivity is usually observed in simple ferromagnetic metals (except Gd), multilayers and composite materials [18]. This is because that the longitudinal resistivity increases monotonically with temperature in most of the metallic materials and a power law dependence of AHE on the longitudinal resistivity. Fe/Gd bilayer is a metallic material and shows a monotonic temperature dependence of longitudinal resistivity (fig. 2c). Therefore a monotonic temperature dependence of AHE should be expected on a Fe/Gd bilayer. The similar features in AHE were also been observed in the single crystal films of SrRuO₃ by Fang et al.. This observation was claimed to be the evidence of the magnetic monopole in the crystal momentum space [10].

To understand the AHE mechanism in a Fe/Gd bilayer we measured the field-dependent Hall resistivity and longitudinal resistivity simultaneously on the patterned samples at different temperatures. From the resulting curves, the temperature dependence of longitudinal resistivity under a zero field and a 5T magnetic field can be easily obtained. The symbols in fig. 2 are the data from the field-dependent resistivity measurements. The lines show the zero field resistivity data for Fe and Gd single layers and Fe/Gd bilayers measured by sweeping the temperature from 5 K to 350 K. It is clearly seen that the resistivities of the three samples exhibit typical metallic behaviors, i.e., the resistivity increases monotonically with temperature. From the resistivity data, it is easy to calculate the magnetoresistance (MR) for the three types of samples as shown in the insets from $MR = (\rho(5T) - \rho(0))/\rho(0)$. It is evident that the MR in monolayer Fe is very small and negligible in comparison with that in monolayer Gd and bilayers of Fe/Gd. This is because, at temperatures far below the Curie temperature, the magnetic spins are well ordered in crystalline iron. Consequently, the magnetic field has a very small effect on the longitudinal electrical resistivity. The temperature dependence of MR in single layer of Gd is quite interesting; a detailed investigation is needed to describe the mechanism.

Shown in figs. 3, 4 and 5 are representative data demonstrating field-dependent Hall resistivity in Fe, Gd and Fe/Gd samples at different temperatures. As expected, the AHE in the single layers of the Fe (fig. 3) and Gd (fig. 4) has very similar behaviors to that of field-dependent magnetization if only magnitude is considered. For thin films of soft ferromagnetic materials, such as Fe, and Gd, the magnetization prefers to lie in the film plan due to the strong de-magnetizing effect. When a magnetic field is applied perpendicular to the film plane, the magnetization is forced to rotate to the field direction. Therefore, the magnetization in the field direction is proportional to the field strength until to the magnetization is fully aligned in the field direction at about 18 kOe and 20 kOe for Fe and Gd film respectively. This process leads to a linearly dependence of magnetization on magnetic field in low field range and a saturation in high field range. Due to the liner dependence of AH resistivity on magnetization (eq. 1), a steep change in AH resistivity is observed at low fields and saturation at high fields. It should be noted that the AHE is positive in Fe layers and increases monotonically with temperature, whereas it is negative in Gd and shows non-monotonic temperature dependence. The temperature dependence of AHE in Fe layers agrees with AHE theories, i.e. AHE increases with increasing longitudinal resistivity. Actually, the intrinsic AHE theories also predict a quadratic dependence [19]. The positive and negative AHEs in Fe and Gd are actually due to the different signs of the spin polarization of the conduction electrons in Fe and Gd, respectively.

To present the data clearly, we plotted the data obtained from Gd at different temperatures in three figures (fig. 4a-c). At temperatures below 150K, the size of the AHE increases with temperature, which can be explained either by the extrinsic model or the intrinsic model. At temperatures above 150K, the AHE deceases. The non–monotonic temperature dependence of AHE cannot be easily understood within the framework of the extrinsic model. The sharp reduction of AHE in Gd with increasing temperature should be due to the sharp decrease in the magnetization as the temperature approaches the Curie temperature of Gd at $T_c \sim 293$ K [20]. The Curie temperature for the Gd film could be lower than that of bulk Gd crystals [20].

Shown in fig. 5 are data showing field-dependent AH resistivity obtained at different temperatures on the Fe (200 nm)/Gd (200 nm) bilayer. As the temperature increases from 5 K to 180 K, the size of AH resistivity increases slightly and reaches its maximum at 180 K. Above 180 K, the magnitude of AHE deceases sharply with increasing temperature and vanishes at about 280 K. Interestingly, above 280 K, the AHE increases again as the temperature increases, but with a positive sign . The behavior of AHE below 280 K is dominated by the AHE in the Gd layer, because the size of the AHE in the Gd film is much larger than that in the Fe film. When the AHE in the Gd layer is comparable with that in the Fe layer, the behavior of the field-dependent AH resistivity becomes quite complex, e.g., consider the curves from 240K to 330K. It is surprising to observe this complex AHE behavior in a bilayer of simple ferromagnetic materials. We believe that these features can be understood as the extension of the model for a Cu/Fe bilayer [17], rather than to interpret them with much more fancy physics models. As shown in fig. 6a, the total current, *I*, along the longitudinal direction flows through both the Fe and Gd layers. Due to the different resistivities of Fe and Gd layers, the current density in the two layers must be different. The AHE of each layer must therefore be calculated using the individual current density. To calculate the current in each layer, we consider the bilayer as two resistors in parallel [21,22,23] (see fig. 6a). Assuming that the resistivity and thickness of each layer are $\rho_{Fe}, \rho_{Gd}, t_{Fe}, t_{Gd}$, the longitudinal current through each layer is then:

$$I_{Fe} = \frac{1}{\frac{\rho_{Fe} t_{Gd}}{\rho_{Gd} t_{Fe}} + 1} I \text{ and } I_{Gd} = \frac{1}{\frac{\rho_{Gd} t_{Fe}}{\rho_{Fe} t_{Gd}} + 1} I \qquad (5)$$

When a magnetic field is applied perpendicularly to the film plane, the Hall voltages and magnetoresistance will be generated in the Fe and Gd layers, respectively. The values of the Hall voltages in each layer are proportional to the applied current, i.e., $I_{Fe}$ and $I_{Gd}$. Since the Hall voltages in the Fe and Gd layer are different both in value and sign, a continuous and constant transverse current must be generated. To understand the mechanism clearly, we draw the circuit for the transverse current (fig. 6b). Here, each layer can be viewed as a battery with an "electromotive force" equal to the AHE generated in that layer and the resistance along transverse direction of each layer is then the inner resistance of the battery. Since AHEs are constant with given applied longitudinal current and magnetic field, the transverse current must be a constant. Assuming that the anomalous Hall resistivities of Fe and Gd are $\rho_{xy,Fe}^{A}$ and $\rho_{xy,Gd}^{A}$, the AHE voltage generated by each layer will be

$$V_{AHE,Fe} = I_{Fe}\rho^A_{xy,Fe}/t_{Fe} \text{ and } V_{AHE,Gd} = I_{Gd}\rho^A_{xy,Gd}/t_{Gd} \tag{6}$$

According to Kirchhoff's Law, the measured total anomalous Hall resistivity of the bilayer is:

$$\rho^A_{xy,bilayer} = \frac{V}{I}(t_{Fe}+t_{Gd}) = \frac{\frac{\rho^A_{xy,Fe}(t_{Gd}/t_{Fe}+1)}{(\rho_{Fe}/\rho_{Gd})(t_{Gd}/t_{Fe})+1} + \frac{\rho^A_{xy,Gd}(t_{Fe}/t_{Gd}+1)}{(\rho_{Gd}/\rho_{Fe})(t_{Fe}/t_{Gd})+1} \cdot \frac{\rho_{Fe}}{\rho_{Gd}}(t_{Gd}/t_{Fe})}{1+\frac{\rho_{Fe}}{\rho_{Gd}}(t_{Gd}/t_{Fe})}. \tag{7}$$

Our Fe/Gd bilayer is a special case of the above model, which has a thickness ratio of $t_{Fe}/t_{Gd}=1$. Hence, from eq. (7), the AH resistivity of our Fe/Gd bilayer is:

$$\rho^A_{xy,Fe/Gd} = 2\left[\frac{\rho^A_{xy,Fe}}{(\rho_{Fe}/\rho_{Gd}+1)^2} + \frac{\rho^A_{xy,Gd}}{(\rho_{Gd}/\rho_{Fe}+1)^2}\right]. \tag{8}$$

Here, we have ignored the magnetoresistance effect, because it has a negligible effect on the Hall voltage.

We calculated the AH resistivity of the Fe/Gd bilayer using eq. (8) and the measured longitudinal and AH resistivities for single layers of Fe and Gd. Shown in fig. 7 are the temperature dependences of the measured AH resistivities at 5 Tesla (symbols) ) for single layers of Fe (200 nm), Gd (200 nm) and the Fe (200 nm)/Gd (200nm) bilayer and calculated AH resistivities (solid line) for the Fe (200 nm)/Gd (200nm) bilayer. As expected, the calculated values are almost identical to the experimental data. This agreement indicates that the proposed model is correct.

Based on the analysis, it is evident that the sign change in the AHE of Fe/Gd bilayer can be interpreted with a simple model without considering the Berry phase in the momentum space. Qualitatively, the sign change is simply due to the competition between two AHE of different signs. Since the size of the AHE for Gd is much larger than that for Fe in the low-temperature range, the behavior of the AHE in the bilayer is dominated by the AHE behavior of the Gd layer. With increasing temperature, the magnetization of the Gd layer decreases and finally becomes very small at its Curie temperature or above. According to eq. 1, the AHE should decrease proportionally with the magnetization, if Rs is independent of the temperature. At or above the Curie temperature of Gd, the AHE for Gd becomes very weak in comparison to that in Fe. Consequently, the sign of the AHE in the bilayer changes to positive.

Non-monotonic temperature and the magnetization dependence of anomalous Hall resistivity, including a sign change, were observed in a bilayer composed of two simple ferromagnetic materials, Fe and Gd. This interesting observation can be understood by considering the competition between the positive and negative Hall voltages generated in the Fe and Gd layers.

The physical picture described by fig. 6 may also suggest that the scaling laws between longitudinal resistivity and anomalous Hall resistivity described by eqs. 2 and 3 cannot be applied to multilayer systems, because the measured AHE and the longitudinal resistivity do not arise from same origin. A sample with a suitable thickness of two ferromagnetic layers with positive and negative AHE that cancels the AHE of the multilayer/bilayer in certain temperature range can easily be prepared. In multilayer or bilayer systems, $\rho_{xy}$ and $\rho_{xx}$ are not simply related to each other by scattering processes because of shunting and transverse current effects.

**Acknowledgements**

We thank Professor **K. W. Wong** for valuable discussions on Berry phase theory. The work described in this paper was supported by grants from the Research Grants Council of the Hong Kong Special Administrative Region (Project No. 604407) and the National Science Foundation of China (No. 50729101 and 50525101).

**Figure Captions:**

Fig. 1. Temperature-dependent saturated Hall resistivity for the Fe/Gd bilayer.

Fig. 2. Temperature-dependent resistivity under zero and 5 T magnetic fields for (a) Fe single layer; (b) Gd single layer and (c) Fe/Gd bilayer. The insets are the temperature dependence of the magnetoresistance measured at different temperatures.

Fig. 3. Representative field-dependent Hall resistivity curves measured at different temperatures on the Fe single layer.

Fig. 4. Representative field-dependent Hall resistivity curves measured at different temperatures on the Gd single layer. (a) 5K to 150 K; (b) 150 K to 270 K; (c) 295 K to 350 K.

Fig. 5. Representative field-dependent Hall resistivity curves measured at different temperatures on the Fe/Gd bilayer. (a) 5K to 150 K (only positive field data displayed for resolvability); (b) 180 K to 350 K.

Fig. 6. Schematic circuit diagram along both directions of the Fe/Gd bilayer. (a) Shunting of the total current along the longitudinal direction (upper) and its equivalent schematic circuit diagram. (b) AHE in the transverse direction (looking along the direction of the applied current) and its schematic circuit diagram (lower).

Fig. 7. Temperature-dependent saturated Hall resistivity for the Fe and Gd single layers and the Fe/Gd bilayer.

**Figures:**

**Fig. 1**

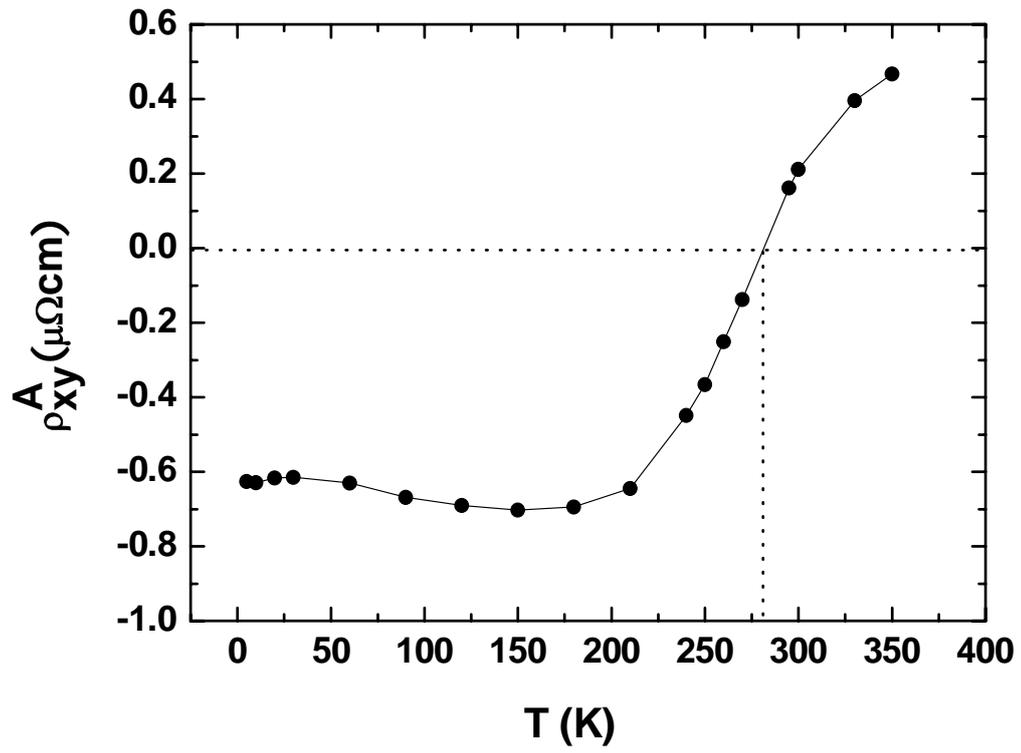

**Fig. 2a**

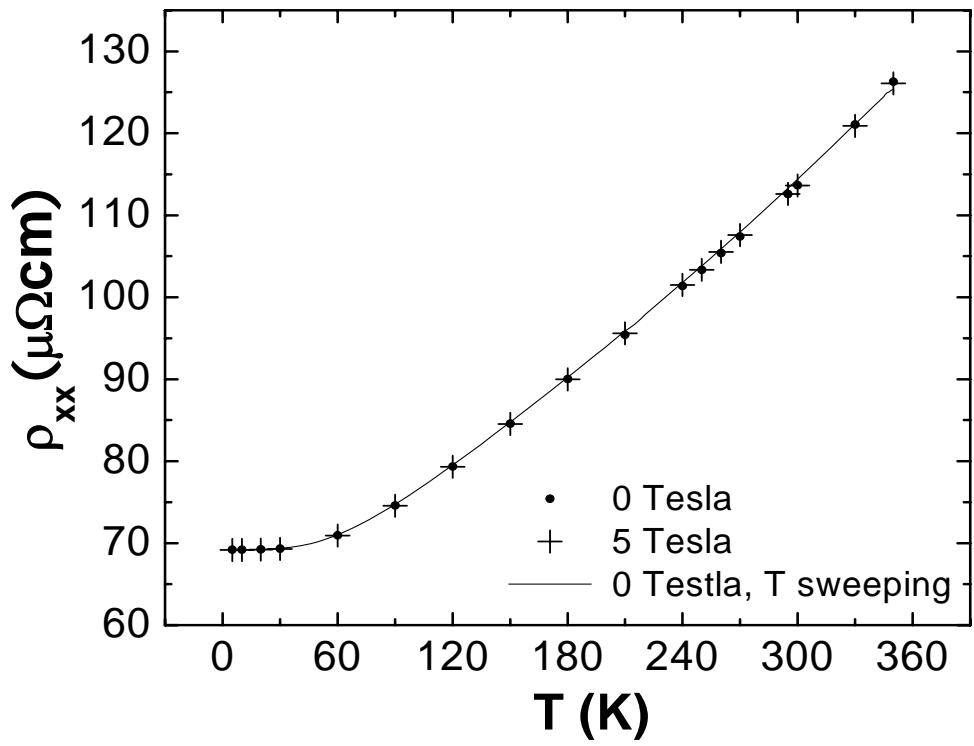

**Fig. 2b**

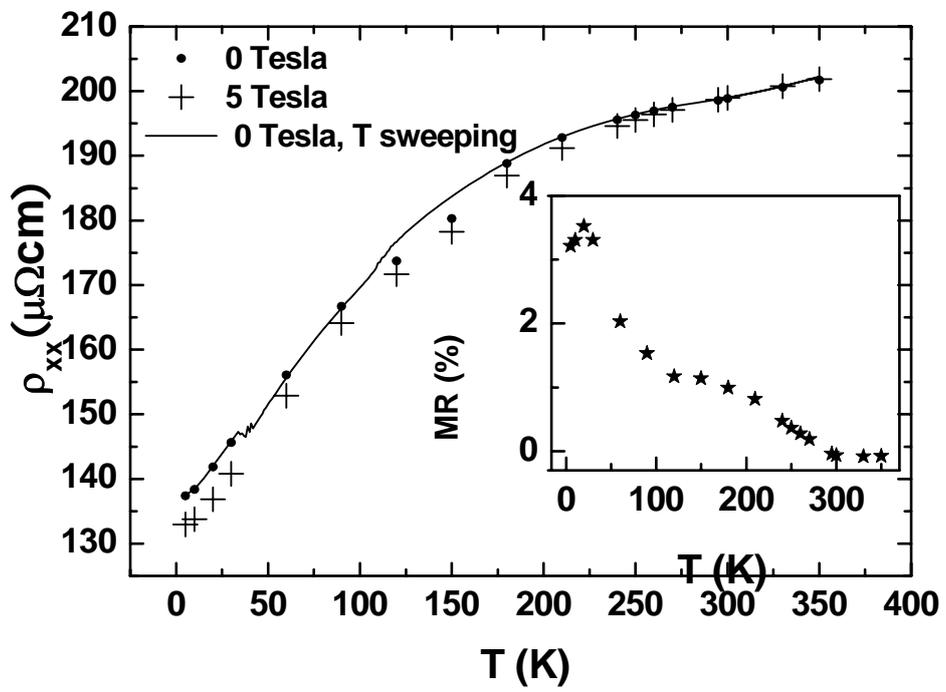

**Fig. 2c**

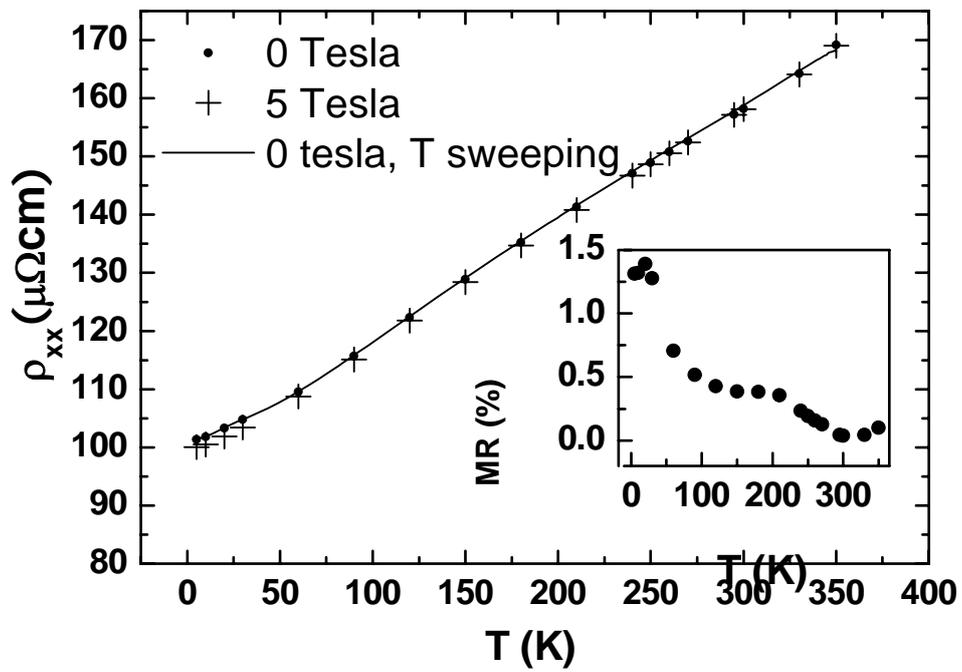

**Fig. 3**

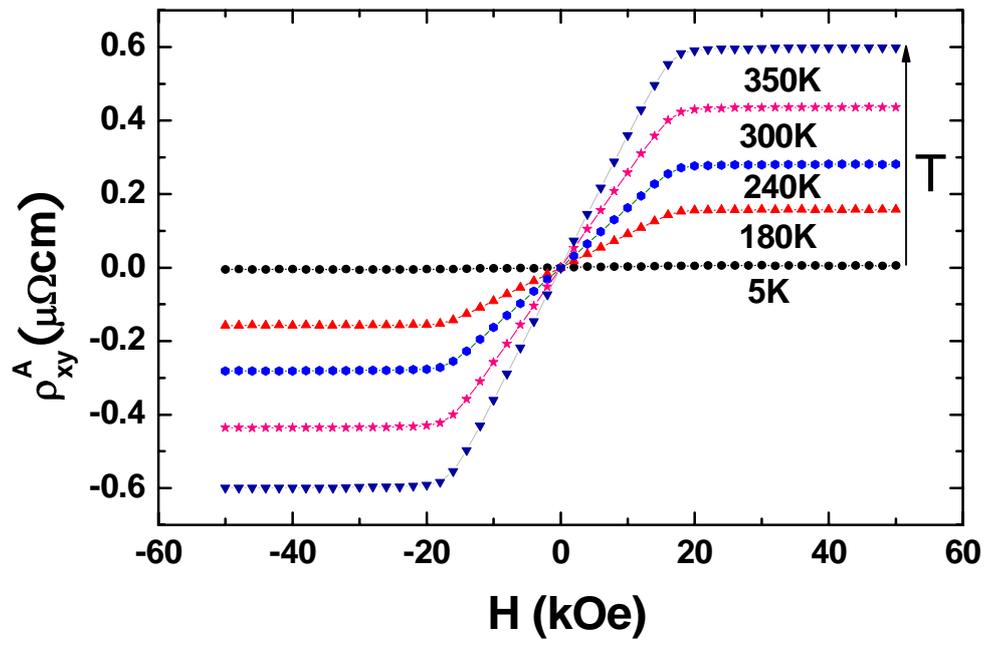

**Fig.4a**

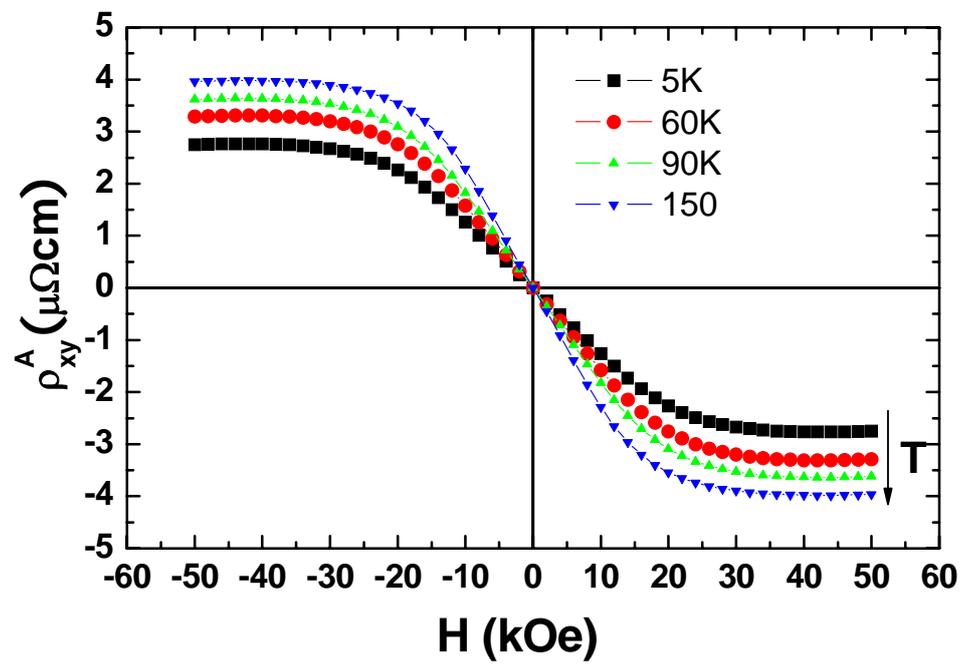

**Fig.4b**

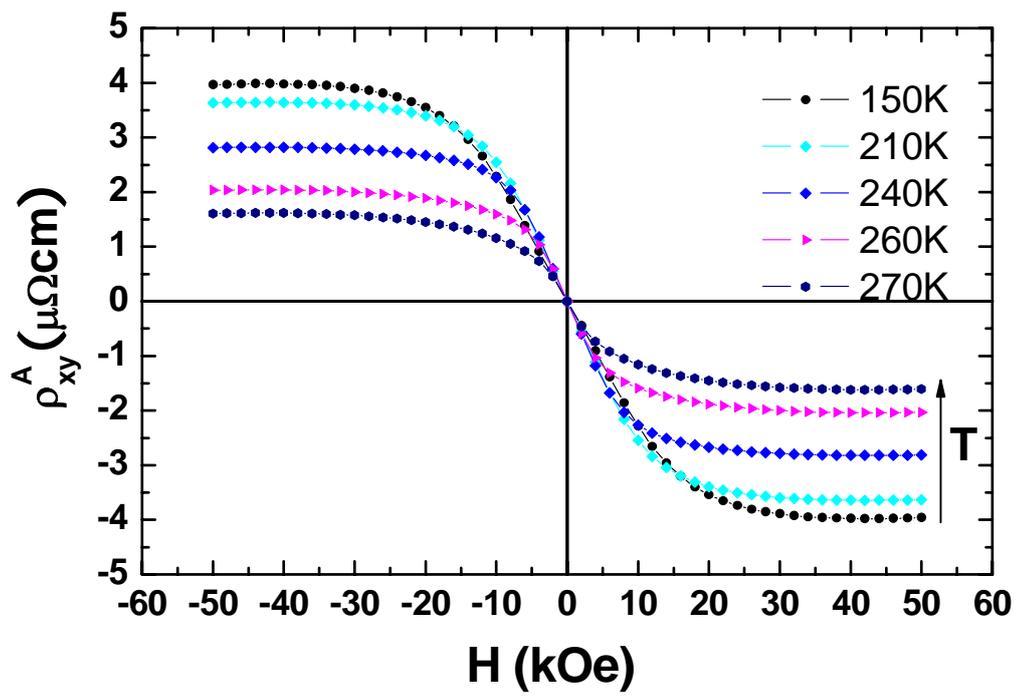

**Fig.4c**

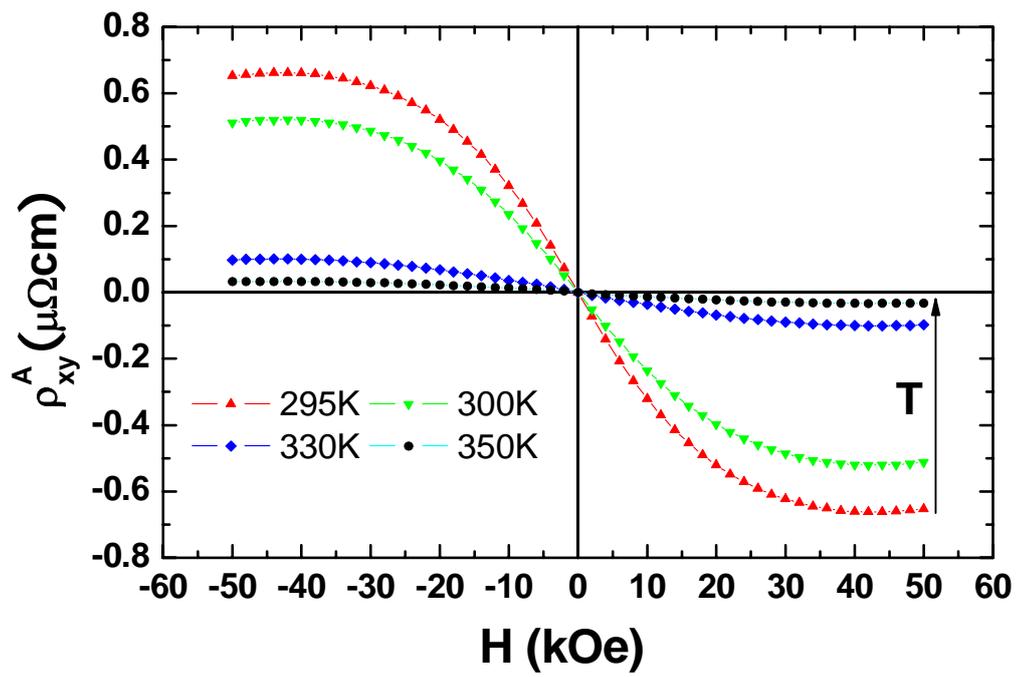

**Fig.5a**

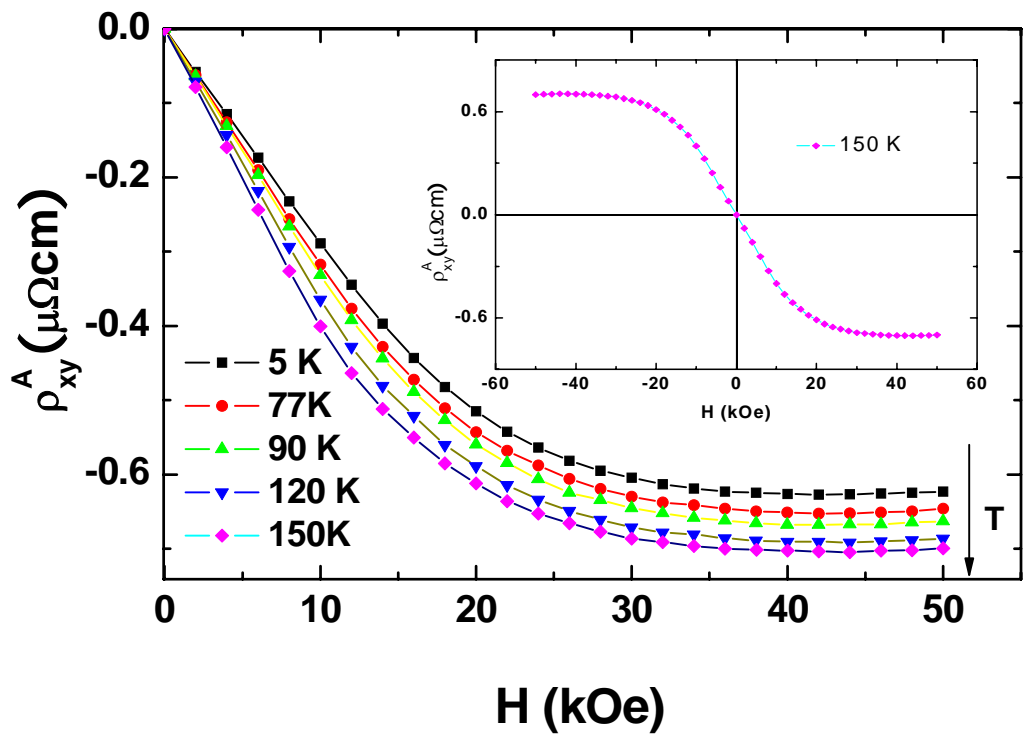

**Fig.5b**

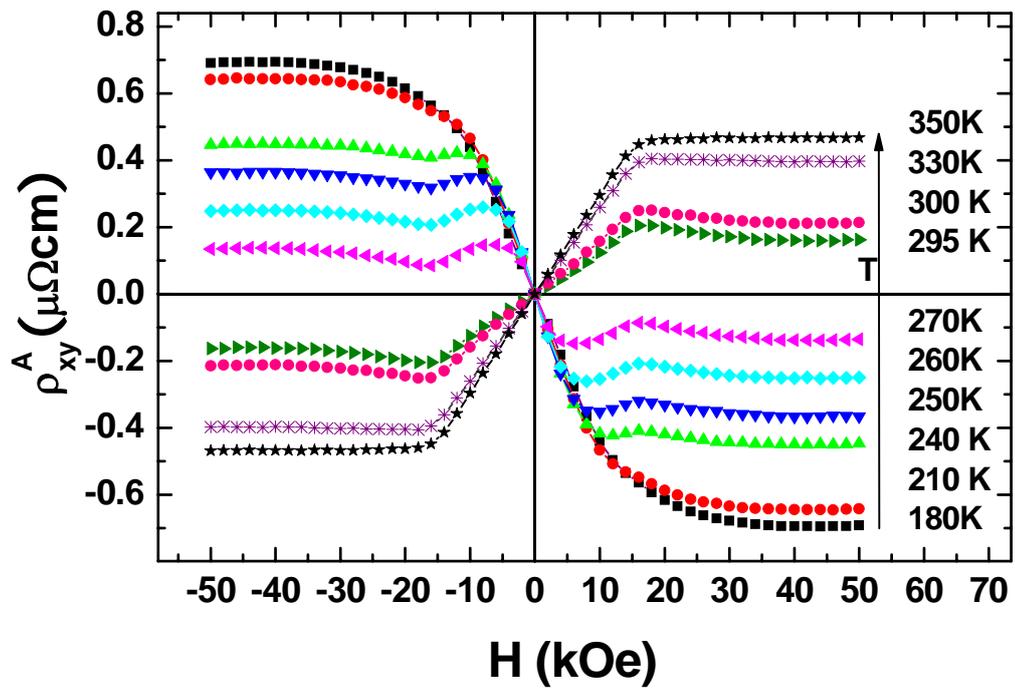

**Fig.6**

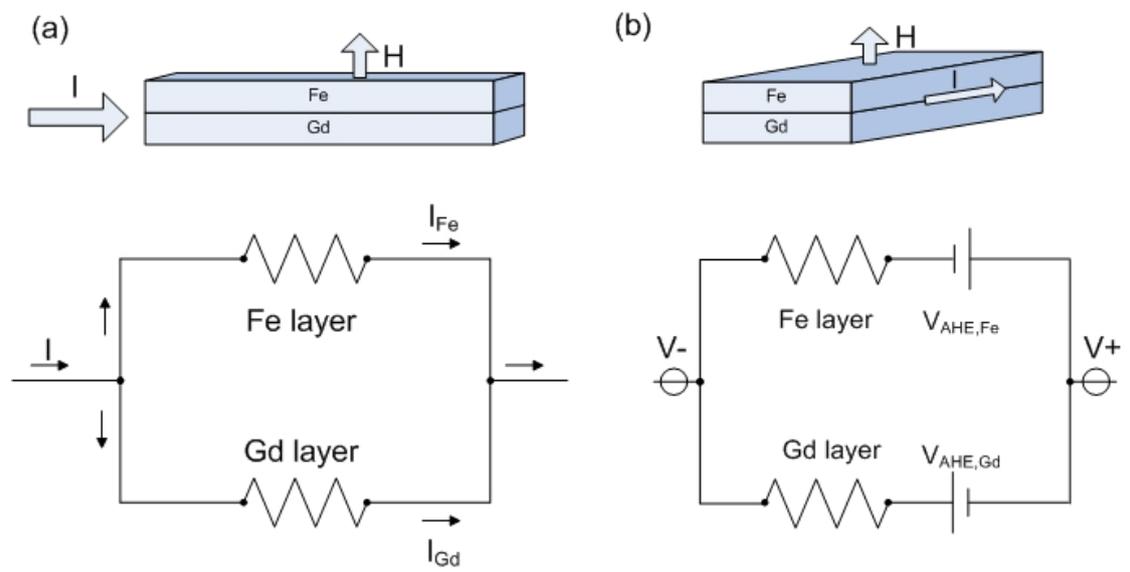

**Fig.7**

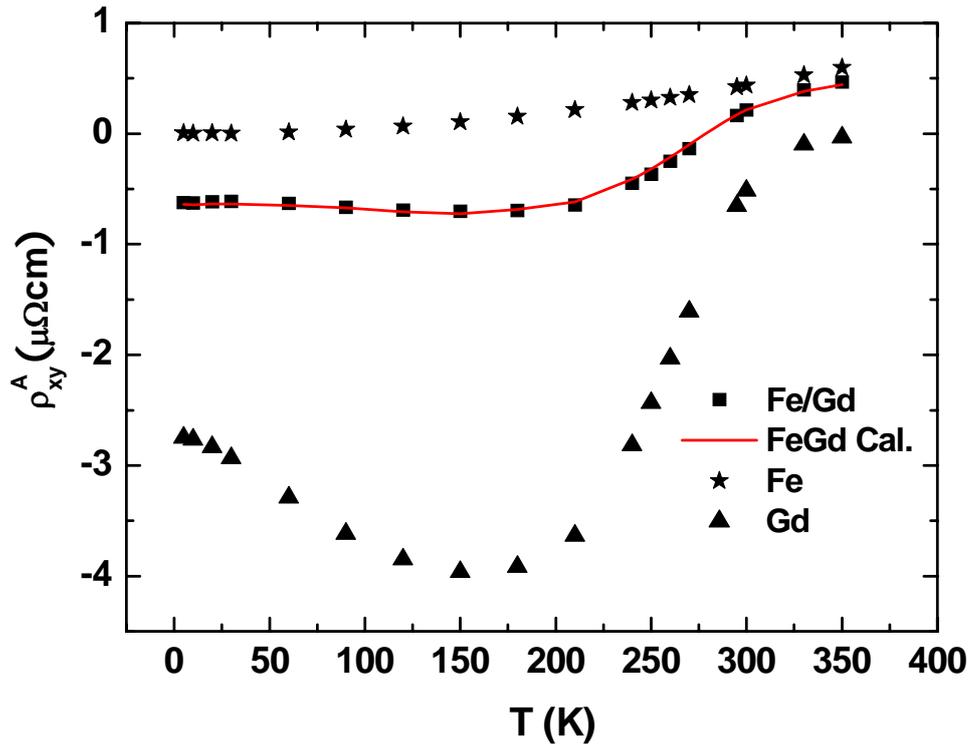